\documentclass[useAMS,usenatbib]{mn2e}

\usepackage{amsmath}
\usepackage{comment}
\usepackage{graphicx}
\usepackage{rotating}
\usepackage{color}
\usepackage{aas_macros}

\newcommand{\beq}{\begin{equation}}
\newcommand{\enq}{\end{equation}}
\newcommand{\beqa}{\begin{eqnarray}}
\newcommand{\enqa}{\end{eqnarray}}
\newcommand{\beit}{\begin{itemize}}
\newcommand{\enit}{\end{itemize}}
\newcommand{\bem}{\begin{pmatrix}}
\newcommand{\enm}{\end{pmatrix}}

\newcommand{\lat}{\left\langle}
\newcommand{\rat}{\right\rangle}
\newcommand{\av}[1]{\lat #1 \rat}

\newcommand{\lb}{\left [}
\newcommand{\rb}{\right ]}
\newcommand{\lp}{\left (}
\newcommand{\rp}{\right )}

\renewcommand{\max}{\mathrm{max}}
\renewcommand{\min}{\mathrm{min}}

\newcommand{\bes}{\begin{sideways}}
\newcommand{\ees}{\end{sideways}}

\newcommand{\EGR}{W}
\newcommand{\yr}{y(r)}
\newcommand{\y}{y}

\newcommand{\yyr}{ y^2(r)}

\newcommand{\rvir}{r_{\textrm{vir}}}

\newcommand{\ephi}{\epsilon_{\Phi}}
\newcommand{\NFW}{\textrm{NFW}}

\title[Statistical ensembles of virialized haloes]{Statistical ensembles of virialized halo matter density profiles}
\author[Carron and Szapudi]{J. Carron\thanks{E-mail:
carron@ifa.hawaii.edu} and I. Szapudi  \\
Institute for Astronomy, University of Hawaii, 2680 Woodlawn Drive, Honolulu, HI, 96822}
\begin{document}

\date{\today}

\pagerange{\pageref{firstpage}--\pageref{lastpage}} \pubyear{2012}

\maketitle

\label{firstpage}

\begin{abstract}
We define and study statistical ensembles of matter density profiles describing spherically symmetric, virialized dark matter haloes of finite extent with a given mass and total gravitational potential energy. 
Our ensembles include spatial degrees of freedom only, a microstate being a spherically symmetric matter density function.  We provide an exact solution for the grand canonical partition functional, and show its equivalence to that of the microcanonical ensemble. We obtain analytically the mean profiles that correspond to an overwhelming majority of microstates. All such profiles have an infinitely deep potential well, with the singular isothermal sphere arising in the infinite temperature limit. Systems with virial radius larger than gravitational radius exhibit a localization of a finite fraction of the energy in the very center. The universal logarithmic inner slope of unity of the NFW haloes is predicted at any mass and energy if an upper bound is set to the maximal depth of the potential well. In this case, the statistically favored mean profiles compare well to the NFW profiles. For very massive haloes the agreement becomes exact.
\end{abstract}

\section{Preliminaries}
Numerical simulations of structure formation in the current successful cosmological paradigm of cold dark matter exhibit long-lived virialized structures, dark matter haloes. Their spherically averaged density profiles as function of radius are well described by a single parameter family of profiles (NFW) \citep{1996ApJ...462..563N,1996grdy.conf..121W,1997ApJ...490..493N,1999ApJ...517...64H,2009MNRAS.396..709W}.  See \cite{2012AnP...524..507F} for a recent review and more references. Moreover, these profiles are universal with respect to the initial conditions of the simulations, yet, it is still unclear to what extent this universality can be explained from first principles. In particular, the interplay and relative importance of statistics and gravitational physics are not fully understood. Motivated by these results and unsolved issues, we pose a relatively simple problem: given only the mass, total gravitational potential energy and spherical symmetry of an ensemble of virialized haloes, 
we seek to find the functional form of the most likely density profiles. 
\newline
\newline   
The present work is not the first attempt at the description of self-gravitating astrophysical systems with statistical arguments. Such ideas started with violent relaxation and the classic paper \cite{1967MNRAS.136..101L} (LB), where the phase space distribution $f(x,v)$ maximizing Shannon's entropy at a given energy is derived \citep[see also][for a recent extension of that approach directed to similar aims]{2013MNRAS.tmp..561P}. The resulting phase space distribution is that of an isothermal sphere, that does not reproduce properly the profiles found in the simulations.
Unlike LB, the present paper is not concerned with the phase space distribution of a single halo, but studies statistical ensembles of haloes. Next we discuss our perspective and its connection to LB.
\newline
\newline
Imagine an ensemble of approximately spherically symmetric, virialized haloes consisting of $N$ particles of some mass $m$, and of a finite extent $\rvir$, the virial radius. If all we are interested in are the spatial degrees of freedom, the complete statistical description of this ensemble is provided by the joint probability density $p_N(x_1,\cdots,x_N)$ of observing particle $1$ at $x_1$, particle $2$ at $x_2$, etc. According to the virial theorem \citep{2008gady.book.....B}, the total energy of each halo is half its gravitational potential energy $W$. A tentative description of the halo ensemble may be for example through a canonical ensemble
\beq \label{Zdiscrete}
p_N(x_1,\cdots,x_N) \propto \exp\lp- \beta \frac G2 m^2 \sum_{i \ne j} \frac{1}{|x_i - x_j|}  \rp.
\enq
The function $p_N$ is the distribution corresponding to maximal entropy for the prescribed mean value of the gravitational energy. The study of the associated partition function and typical realizations does not appear simple in this discrete formulation.  Neither it is obvious how to enforce spherical symmetry in \eqref{Zdiscrete}. The principal purpose of this paper is to solve exactly the statistical ensembles analog to \eqref{Zdiscrete}, where the probability density $p_N(x_1,\cdots,x_N)$ is replaced by a probability functional $p[\rho(x)]$, and $\rho(x)$ is a spherically symmetric positive function.
\newline
\newline
The non-consideration of the velocity degrees of freedom and thus of the kinetic energy is far from innocent. Together with the presence of a sharp boundary $\rvir$, it is in fact one of the elements making our statistical ensembles well defined and leading to meaningful results. It is indeed well known that the thermodynamics of gravity is intrinsically difficult \citep{1968MNRAS.138..495L,1999PhyA..263..293L,2008gady.book.....B}, and in particular the statistical ensembles are often not well defined due to the unbounded available phase space. Identifying the energy of the halo with its gravitational energy with through the virial theorem short-cuts many such difficulties. On the other hand, the specific thermodynamics associated with our ensembles are generally different from that of gravity.   
\newline
\newline
The connection to LB can be made apparent with the following argument. The probability  $p_1(x)$ to observe a particle  (particle $1$, or any other) at $x$ is obtained by integrating out the $N-1$ other coordinates of $p_N$. We enforce spherical symmetry by enforcing the same on $p_1$. A realization of the density within the ensemble is $\rho(x) = \sum_{i}\delta^D(x - x_i)$, and the mean density profile is directly proportional to $p_1$,
\beq
\av{ \rho}(x) = N p_1(x).
\enq
Now, let us assume hypothetically that the following uncorrelated form of the $N$ particle probability density holds
\beq
p_N(x_1,\cdots,x_N) = \prod_{i = 1}^N p(x_i).
\enq
The entropy of $p_N$ is now simply $N$ times that of $p_1$. Maximizing the entropy with fixed mean gravitational potential energy leads now to
\beq \label{LB}
p_1(x) \propto \av{\rho}(x) \propto e^{\beta m \av{\Phi}(x)}.
\enq
This is identical to the density profile obtained by LB, isothermal spheres with profiles decaying as $\rho(r) \propto r^{-2}$ at large radii. In this equation, $\av{\Phi}(x)$ is the gravitational potential at $x$, obtained consistently from the mass distribution $\av{\rho}(x)$. In the LB approach, $\beta$ is the inverse temperature of the velocity distribution, a Maxwell-Boltzmann distribution. 
\newline
\newline
Thus, one interpretation of the approach taken in this paper, especially for the grand canonical ensemble, is the LB entropy maximization approach to the $N$-particle distribution in a continuous description, ignoring the velocity degrees of freedom. Indeed, our approach recovers the singular isothermal spheres in a particular regime, but also interesting new phenomena arise such as localization of energy, or the possibility of negative temperatures.
\newline
\newline
This paper is organized as follows. In section \ref{sectionmacro} we solve the statistical ensembles associated to our problem. More specifically, we first find the partition function of the grand canonical ensemble in section \ref{sectiongd}, and then discuss in \ref{sectionmicro} its equivalence to the microcanonical ensemble in the relevant limit. We obtain the mean density profiles that can be realized in the most microstates. We then compare these profiles to NFW profiles in section \ref{sectionNFW}. We conclude with a discussion in section \ref{sectionconclusion}.
\section{Solving the statistical ensembles} \label{sectionmacro}
For a spherically symmetric halo of extent $\rvir $ with density profile $\rho(r)$, the mass and gravitational energy are given by
\beq
M[\rho] = 4\pi \int_0^{\rvir} dr\:r^2 \rho(r),
\enq
and
\beq
W[\rho] = -\frac G 2 \int_0^\infty dr \: \lp \frac{M(\le r)}{r} \rp^2,
\enq
where $M(\le r)$ is the mass enclosed within $r$.
The microcanonical ensemble, expressing uniform statistical weight on the surface of constant mass and energy, is formally \beq
\Omega(M,W) \propto \int \mathcal D\rho \: \delta^D(M - M[\rho])\delta^D(W - W[\rho])
\enq
where the integration runs over all positive functions $\rho(r)$ in a sense that we will make precise later, and $\delta^D$ is the Dirac delta function. Similarly, the grand canonical ensemble reads as 
\beq
Z(\lambda,\beta) \propto \int \mathcal D\rho \:e^{-\lambda M[\rho] - \beta |W[\rho]|}.
\enq
We present a solution for the latter first, establishing their equivalence afterwards. The parameter $\beta$ can be interpreted as an inverse temperature $1/k_B T$, and the parameter $\lambda$ as $-\mu / k_B T$, where $\mu$ is the chemical potential. Equivalently, according to Jaynes' \citep{jaynes57,jaynes83} perspective on statistical mechanics,  the corresponding probability density $p[\rho]  \propto e^{-\lambda M - \beta |W|}$ is the probability density of maximal entropy at a given mean mass and energy.
\newline
\newline
Before calculating the partition function, let us introduce the following function of radius
\beq \label{defyr}
\yr := 4\pi \int_{r}^{\rvir} ds\:s\rho(s)
\enq
This function is the most fundamental for our purposes, as it will
allow the factorization of the partition function. Another 
representation of $\yr$ is
\beq \label{potential}
\Phi(r) = -\frac G r \int_0^r ds\: y(s),
\enq
where $\Phi(r)$ is the gravitational potential at $r$, thus the average of $\yr$ between $0$ and $r$.
From its very definition the $\yr$ function is always a positive, decreasing function of radius until it reaches zero at the virial radius. Further, the maximal value $\y(0)$ that the $\y$-function takes is the central potential of the mass distribution. From equation \eqref{potential} 
\beq
\Phi(0) =  -G\:\y(0).
\enq
Note that both sides of that equation can be infinite.
\newline
\newline
The $\y$-function has the following two convenient relations: for any spherically symmetric distribution holds that the mass and energy are its first and second moments:
\beq \label{mass}
M = \int_0^{\rvir} dr \: \yr
\enq 
as well as
\beq \label{energy}
\EGR = -\frac{G}{2} \int_0^{\rvir} dr\:\yyr.
\enq
A proof of these two identities can be found in the appendix.
\newline
\newline
The relevant, dimensionless parameter in this work is the ratio $a$ of the virial radius to the gravitational radius
\beq
a := \frac{\rvir}{r_g},
\enq
where
\beq
r_g = \frac{GM^2}{|W|}.
\enq
For future use, we introduce the depth of the central potential well as another parameter
\beq \label{ephi}
\frac 1 {\ephi} := \frac{\Phi(0)}{\Phi(\rvir)} = y(0)\lp  \frac{\rvir}{M}\rp.
\enq
One can show from equations \eqref{mass} and \eqref{energy} that any mass distribution must have
\beq
\frac 12 \le a \le \frac{1}{2\ephi}.
\enq
The singular isothermal sphere $\rho \propto r^{-2}$ has $a = 1, \ephi = 0$ and plays a special role in this work. 
\subsection{The grand canonical ensemble} \label{sectiongd}
\begin{figure}
\begin{center}
\includegraphics[width = 0.45\textwidth]{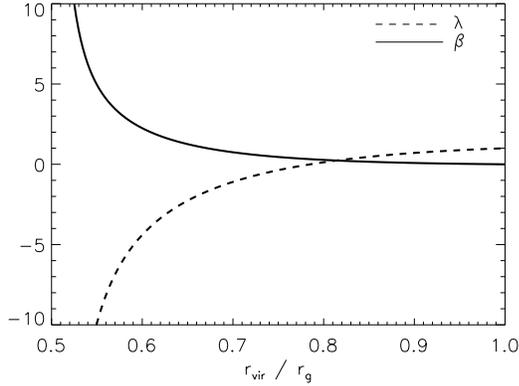}
\caption{{The dimensionless parameters $\lambda$ and $\beta$ of the grand canonical ensemble as function of $\rvir / r_g = a$. These parameters are set by equations \eqref{eq1} and \eqref{eq2}}. These equations have no solution for $a > 1$. This regime is better understood with the microcanonical ensemble.}
\label{figpotentials}
\end{center}
\end{figure}

\begin{figure}
\begin{center}
\includegraphics[width = 0.45\textwidth]{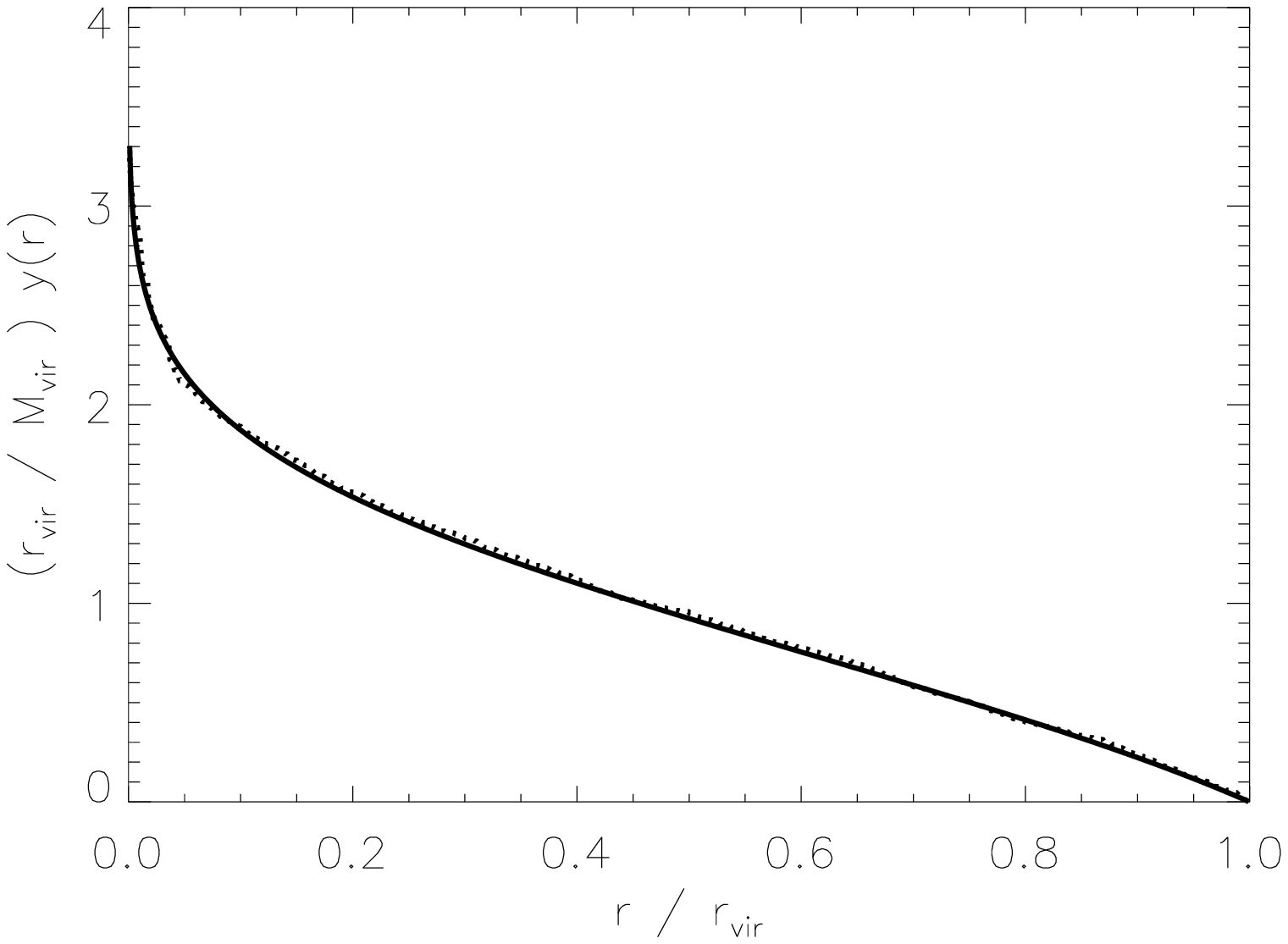}
\includegraphics[width = 0.45\textwidth]{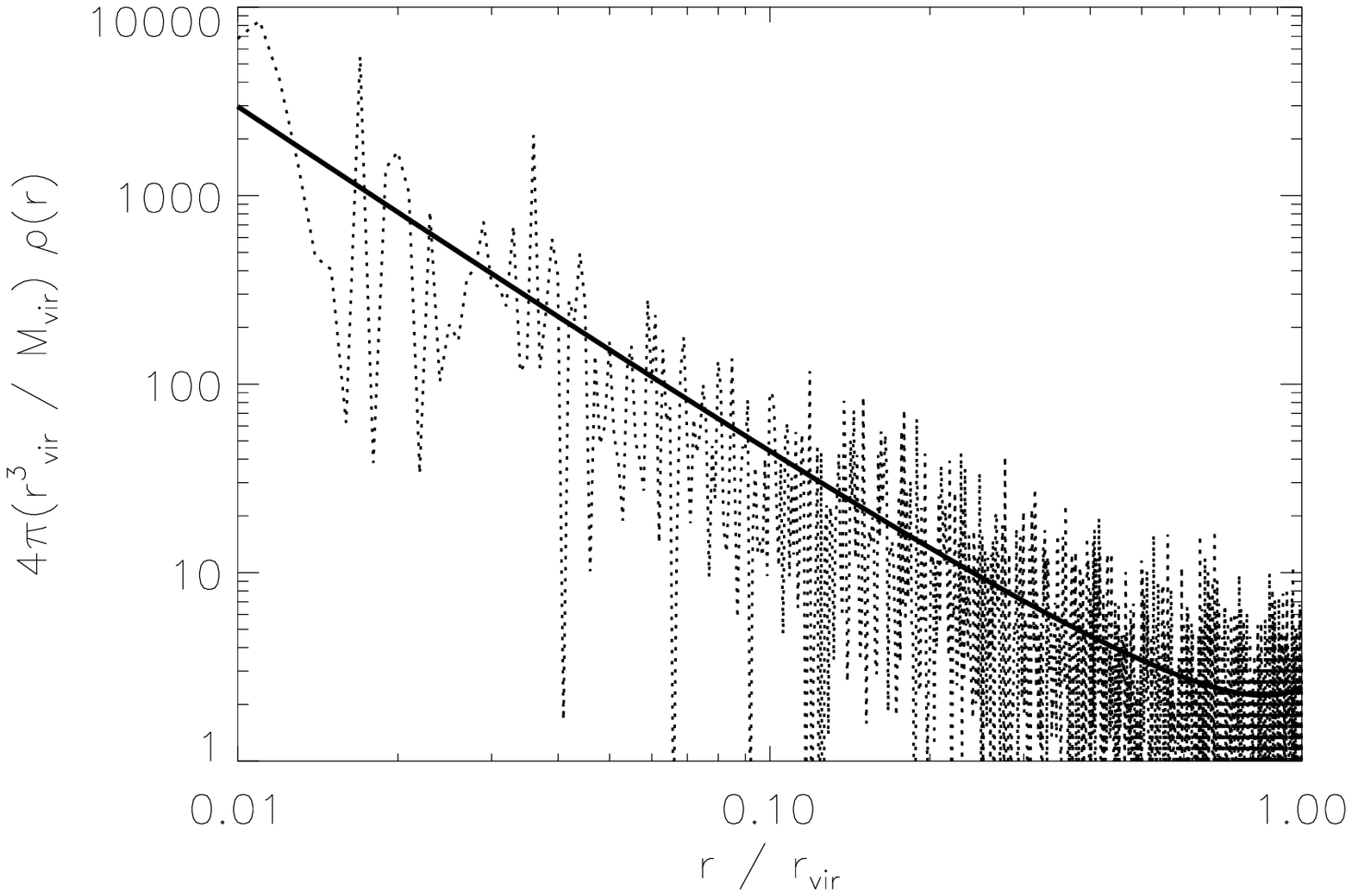}
\caption{A realization of a microstate of the grand canonical ensemble \eqref{ensemble}(dotted lines), for $a  = 0.7$ and $N = 1000$, together with their exact mean value (solid lines).  The upper panel shows the dimensionless $y$-function. The lower panel displays the dimensionless density profile $u(x)$. For large $N$, the ensemble always makes a sharp prediction on the $y$-function and other integrated quantities, but realizations of the density profiles themselves fluctuate around the mean profile.}
\label{figy}
\end{center}
\end{figure}

In terms of the $\y$-function, the partition function of the grand canonical ensemble is given by
\beq \label{Zy}
Z\lp \lambda,\beta\rp = \int \mathcal D \y(r) e^{ -\lambda \int_0^{\rvir} dr\:y(r)  - \beta \frac{G}{2}\int_0^{\rvir}dr\:y^2(r) },
\enq
with boundary conditions
\beq
 y(\rvir) = 0,  \quad y(r) \ge 0,\textrm{  and}\quad y'(r) \le 0.
\enq
The Jacobian of the transformation from $\rho$ to $y$ is of no relevance, since the relation is linear.
The parameters $\beta$ and $\lambda$ are related to the mean energy and mass by the two non-linear equations
\beq
|\EGR| = -\frac{\partial \ln Z}{\partial \beta}
\enq
and
\beq
M = -\frac{\partial \ln Z}{\partial \lambda}. 
\enq
From here on we work with dimensionless functions and potentials $\lambda,\beta$. We define the dimensionless $y$-function
\beq
z(x) = \lp \frac {\rvir}{M} \rp \yr, \quad x = \frac r {\rvir},
\enq
and the dimensionless density profile
\beq \label{ux}
u(x) =- \frac{z'(x)}{x} = 4\pi \lp \frac{\rvir^3}M \rp \rho(r).
\enq
We approach the formal expression \eqref{Zy} through discretization, replacing the continuous function by its sample on a regular grid of $N$ points, replacing integrals by sums, and derivatives by finite differences. We then obtain a solution that has a well defined limit for $N \rightarrow \infty$.
\newline
\newline
We define $z_i := z(x_i), x_i := i/(N+1), \quad i = 1,\cdots,N$, and set further $\int_0^1 dx\: z(x) \approx \frac 1 N\sum_k z_k $ and $\int_0^1 dx \:z^2(x) \approx \frac 1 N \sum_k z^2_k $. After renormalization of the potentials, the partition function is 
\beqa \label{ensemble}
Z_N(\lambda,\beta) &= \int_0^\infty d^Nz \prod_{k = 1}^N \exp \lp -\lambda z_k - \beta z_k^2 \rp \\
&z_1\ge z_2 \ge \cdots \ge z_N.
\enqa
Using the symmetry of the integrand, this is simply
\beq
Z_N(\lambda,\beta) = \frac{1}{N!} \lp \int_0^\infty dz\: \exp \lp -\lambda z - \beta z^2\rp \rp^N \nonumber
\enq
\beq
=: \frac {1}{N!}Z_1(\lambda,\beta)^N.
\enq
The defining equations for $\lambda$ and $\beta$ are seen to become
\beq \label{eq1}
1 = -\frac{\partial \ln Z_1}{\partial \lambda} = \frac{1}{Z_1}\int_0^{\infty} dz\:z\:e^{-\lambda z - \beta z^2} 
\enq
and
\beq \label{eq2}
2a = -\frac{\partial \ln Z_1}{\partial \beta} = \frac{1}{Z_1}\int_0^{\infty} dz\:z^2e^{-\lambda z - \beta z^2}.
\enq
Interestingly, $\lambda$ and $\beta$ are now set by the simple matching of the first two moments of the one-dimensional distribution
\beq \label{pz}
p(z) := \frac {1}{Z_1}\exp\lp -\lambda z - \beta z^2\rp,
\enq
a Gaussian distribution with range truncated to $0 \le z \le \infty$.
\newline
\newline
This ensemble we are dealing with after discretization is in fact very well known from the field of order statistics \citep{1977ats..book.....K,Wilks}. The probability density in equation \eqref{ensemble} is the probability density describing the ordering in decreasing order of $N$ values drawn independently from the same distribution $p(z)$. Microstates $z(x)$ can be generated easily following that procedure, since $z\lp k/(N+1) \rp$ is simply the $k$th largest value of the sample of size $N$. The properties of the haloes generated by this ensemble that we will expose below can therefore be obtained from well known results with no difficulty.\newline
\newline
The two equations \eqref{eq1} and \eqref{eq2} have a solution for $1/2 < a \le 1$, that can easily be obtained with standard numerical methods. The solution is shown on figure \ref{figpotentials}. The lower bound $a = 1/2$ simply expresses the fact that there are no spherically symmetric distribution with $a < 1/2$. The upper bound $a = 1$, corresponds to infinite temperature $\beta = 0$, and shows a transition to a regime that the grand canonical ensemble fails to describe properly. We will solve this regime later in section \ref{sectionmicro}.
\newline
\newline
In particular, the ensemble makes for large $N$ a sharp prediction on the value of $z(x_i)$, fluctuations decaying to zero for large $N$. For $0 \le z \le \infty$, let $F(z)$ be
\beq
F(z) = \frac{1}{Z_1}\int_0^z dy\:e^{-\lambda y - \beta y^2},\quad 0 \le F \le 1 
\enq
and $Q(x)$ be its inverse function
\beq
Q(x) = F^{-1}(x).
\enq
In statistical jargon, $F$ is the cumulative distribution function of $p(z)$ and $Q$ the quantile function. We have that the $z$-function singled out by the ensemble is given by for large $N$
\beq \label{yGC}
z\lp \frac{i}{N+1} \rp \stackrel{N \gg 1}{\rightarrow} Q\lp 1 - \frac{i}{N +1} \rp,\quad i = 1,\cdots,N
\enq
that can be written in the limit of a continuum,
\beq
 z(x) = Q(1-x), \quad 0 < x < 1.
\enq
The corresponding mean density profile is obtained from $z(x)$ following equation \eqref{ux}, with the result
\beq \label{rhoGC}
u(x) = \exp\lp \ln Z_1[\lambda,\beta] + \lambda z(x) + \beta z^2(x) - \ln x \rp.
\enq
The mass and energy of the microstates also become sharply defined for large $N$, since they are obtained integrating $z(x)$. This fact already points to the equivalence of the grand canonical and microcanonical ensembles discussed later in more detail.
\newline
\newline  
On the other hand, the density profile $u(x) = z'(x)/x $ involves at any finite $N$ the finite differences $u_i \propto N(z_{i} - z_{i+1})/x_i$. Fluctuations around the mean profile \eqref{rhoGC} do not decay in the large $N$ limit. Rather, it can be shown that the density profile at different arguments behave like a collection of independent, exponentially distributed variables centered on the mean profile \eqref{rhoGC}. This is clear for instance for the case of $\beta = 0$ where the ensemble $\propto e^{- \lambda M}$ does not introduce correlations between the density at different arguments. This is illustrated on figure \ref{figy}, showing in the upper panel a realization of the function $z(x)$ and in the lower panel that of the the density profile, for $a = 0.7$ and $N = 1000$. The black lines are the exact mean values in the large $N$ limit, equations \eqref{yGC} and \eqref{rhoGC}.
\newline
\newline
For $\beta = 0$, corresponding to the upper bound $a = 1$, we have further through direct inspection $Q(x) = -\ln (1-x)$, so that
\beq \label{zSIS}
z(x) = - \ln \lp  x \rp, \quad \beta = 0
\enq
and we recover thus the profile $\rho \propto r^{-2}$ of the singular isothermal sphere.
 At finite temperature, an exact if not necessarily very insightful form for $z(x)$ is
\newcommand{\erfc}{\textrm{erfc}}
\beq \label{zfunction}
z(x) = \frac{1}{\sqrt {\beta}}\lb \erfc^{-1} \lp x\: \erfc \lp \gamma \rp \rp - \gamma \rb,\quad \beta >0
\enq 
with $\gamma = \lambda / 2 \sqrt \beta$, and where $\erfc(x)$ is the complementary error function. Finally, the lower limit $a = 1/2$ corresponds to $p(z) \approx \delta^D(z - 1)$, leading to the degenerate profile where all the mass is concentrated at the virial radius. The outer slope is found to be
\beq
\left. \frac{d \ln \rho}{d \ln r} \right|_{r = \rvir} = -1 - \sqrt{\pi} \gamma\: e^{\gamma^2}\erfc(\gamma),
\enq
and transitions smoothly from $+\infty$ at $a = 1/2$ to $-2$ at $a = 1$. 
\newline
\newline
We now turn to the more careful study of the regime $a > 1$.
\subsection{The regime $\rvir/r_g > 1$ and the microcanonical ensemble} \label{sectionmicro}
In the discretization introduced above, the microcanonical ensemble takes the following form,
\beqa
\nonumber \Omega_N &= \int d^Nz\:\delta^D\lp 1 - \frac 1 N \sum_{k} z_k \rp\delta^D \lp 2a - \frac 1N \sum_{k} z_k^2\rp \\
&\infty \ge z_1\ge z_2 \ge \cdots \ge z_N \ge 0.
\enqa
By symmetry, it reduces to
\beq
\Omega_N = \frac 1 {N!}\int_{0}^\infty d^Nz\:\delta^D\lp 1 - \frac 1 N \sum_{k} z_k \rp\delta^D \lp 2a - \frac 1N \sum_{k}z_k^2\rp.
\enq
The mass (in our units unity) defines a plane in $N$-dimensional cartesian space, while the energy $2a$ defines a sphere. The microcanonical ensemble is remarkably reduced to the (rather formidable) geometric problem of the intersection of a plane and a sphere in $N$ dimensions, together with the constraints $z_i \ge 0$.  A realization of $z$ is obtained by drawing at random a vector on this set, and then ordering its coordinates. 
\newline
\newline
Unlike for the grand canonical ensemble, the calculation of the $y$-function at fixed $N$ does not appear tractable with simple means. Nonetheless, the uniform distribution on this set (without ordering) has already been studied in the limit of large $N$ \citep{2010arXiv1011.4043C}, and we can use the main results exposed there. Notably, the equivalence to the grand canonical ensemble for $1/2 < a \le 1$ rigorously follows directly from Theorem 1.1 in that work \footnote{This theorem states that for $1/2 < a \le 1$ any finite number of the coordinates of $z$ converges in law to independent identically distributed variables with probability density the Gaussian restricted to the positive axis as given in equation \eqref{pz}. In the notation of \cite{2010arXiv1011.4043C} we have $\lambda = r, \beta = s$ and $2a = b$}.
\newline
\newline
Theorem 1.2 of \cite{2010arXiv1011.4043C}  deals with the case $a >1$, demonstrating localization of energy. It is shown that the largest component of such a random vector $z$ (in our notation, since $x_1 = 1/(N+1)$, this largest component is $z(x_1) \simeq z(1/N)$) grows in such a way with $N$ that the part of energy $\frac 1 N z^2(1/N)$ that it carries is finite with probability 1. It is not a localization of mass in the sense that the fraction of mass $\frac 1N z(1/N)$ still tends to zero. The amount of energy localized in the center is precisely the surplus of energy with respect to the singular isothermal sphere $a = 1$. The second largest coordinate does not show this localization property. The density profile that we recover is in fact exactly that of the singular isothermal sphere with a stronger singularity in the very center $x = 0$ accounting for the additional energy. Note that the situation is reminiscent to Bose condensation.
\newline
\newline
The above results can also be seen from the perspective of a modified grand canonical ensemble. The grand canonical ensemble fails for $a > 1$ because a positive inverse temperature $\beta$ is required, for the system can reach arbitrarily high energies $|W|$. With $\beta < 0$ the probability density $p(z)$ cannot be normalized anymore. As in other physical systems we can allow negative temperatures by setting an upper bound to the energy that the system can have access to. This is simply done by introducing a maximal value for the $z$-function. Since $z(0)$ is the potential well $1/\epsilon_\Phi$ in \eqref{ephi}, this is equivalent to force the central potential of the mass distribution to take a finite value.
\newline
\newline
Our results of the previous section, equations \eqref{ensemble} to \eqref{rhoGC} all hold unchanged provided the range of $z$ in $p(z)$ is now $0 \le z \le 1/\epsilon_\Phi$. In particular, to each $a$ within the allowed range $1/2 < a < 1/2\epsilon_\Phi$ there is a corresponding $\lambda$ and $\beta$. Since $z(x)$ is given by the quantile function $Q(1-x)$ of $p(z)$ we have
\beq
z(0) = \frac{1}{\epsilon_\Phi}.
\enq
The profile singled out by the ensemble always saturates the bound. 
\newline
\newline
To very small $\ephi$ and $a > 1$ correspond now high negative temperatures $\beta <\approx 0$ and $\lambda \approx 1$. This implies that $p(z)$ is over a long range the exponential distribution, with relevant deviations only for $z \approx 1/\ephi$ accounting for the match of the second moment of $p(z)$. Thus we recover as we just argued the exact singular isothermal spheres with the corresponding additional singularity in the center. As we show next, the limit introduced for the central potential has the additional benefit of recovering the universal NFW profiles seen in simulations.
\section{Negative temperature profiles and NFW haloes}\label{sectionNFW}
\begin{figure}
\begin{center}
\includegraphics[width = 0.45\textwidth]{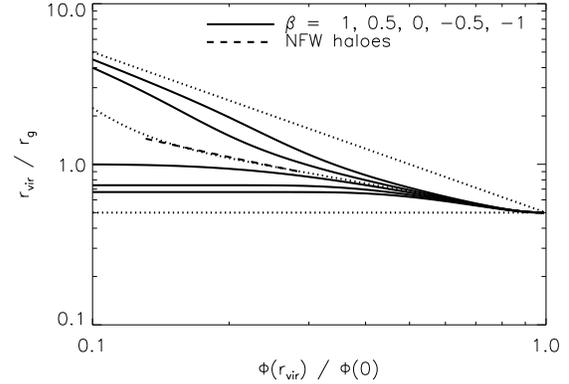}
\caption{The dashed line show the location of the spherically symmetric NFW haloes ($3\le c \le 15$) in the plane set by $a = \rvir/r_g$ and $\ephi = \Phi(\rvir)/\Phi(0)$, with more massive haloes having larger $\ephi$. There is no spherically symmetric mass configuration outside the region delimited by the two dotted lines $a = 1/2$ and $a = 1/2\ephi$. The solid lines illustrate the behavior of the line of constant $\beta$ of the grand canonical ensemble. The NFW haloes are seen to lie close to the line of constant $\beta = -0.17$, the dotted line.}
\label{figNFWc}
\end{center}
\end{figure}

The one-parameter family of NFW profiles is given by
\beq \label{rhoNFW}
u_{\NFW}(x) \propto \frac 1x \frac{1}{\lp 1 + c x \rp^2},
\enq
where $c$ is the concentration parameter of the halo. More massive haloes have smaller concentration, and the range of haloes probed in numerical $N$-body simulations corresponds approximately to the range $3 \simeq c \simeq 15$. The corresponding $y$-function is given by
\beq \label{yNFW}
z_{\NFW}(x) \propto \frac{1}{1 + cx} - \frac {1}{1 + c}.
\enq
From this relation follows
\beq
\epsilon_{\Phi,\NFW} = \frac 1 {c^2} \lb (1 + c)\ln (1 + c) - c \rb,
\enq
as  well as
\beq
a_{\NFW} = \frac 1{2c}\lp \frac{1 -2\epsilon_{\Phi,\NFW} }{\epsilon^2_{\Phi,\NFW}}\rp.
\enq
\newline
\newline
It is clear that the ensembles at positive temperature $\beta > 0$ are unable to accommodate for these profiles. Our study of these ensembles in the previous section showed that the overwhelming majority of profiles at given mass and gravitational potential energy have their $y$-function given by \eqref{zfunction} for $a < 1$, or basically by the singular isothermal sphere \eqref{zSIS} for $a \ge 1$. None of these profiles can provide a satisfactory match. The outer slope at $x  = 1$ is not steep enough, and the singularity in the center is too strong. The NFW  density profile is singular in the center, but as can be seen from \eqref{yNFW} all NFW haloes have a finite central potential $z(0)$, equivalently $\ephi > 0$. The profiles singled out by the ensembles all have $z(0) = \infty, \ephi = 0$.
\newline
\newline
If some process is preventing the growth of the central potential, at least over the relevant time scales, we can still ask whether the ensembles with the additional constraint of a finite potential well introduced in section \ref{sectionmicro} can reproduce these profiles more successfully.
\newline
\newline
There is a strong phenomenological argument supporting this view. As noted previously, the $y$-function of these ensembles are smooth functions that now tend to the finite value $1/\ephi$ at $x = 0$. Therefore, the behavior of the density profile $z'(x) / x$ in the center predicted by these ensembles is precisely the power law of the NFW profile
\beq
z(x) \propto \frac 1x, \quad x \rightarrow 0
\enq
at any mass and energy.
\newline
\newline
The dashed line in Figure \ref{figNFWc} shows the  the location of the NFW haloes with $3 \le c \le 15$ in the $(\ephi -a)$ plane, together with several lines of constant $\beta = 1,0.5,0,-0.5,-1$ (solid lines, from bottom to top). The dotted lines $a = 1/2$ and $a = 1/2\ephi$ corresponds to $\beta = \pm \infty$, or zero temperature, where there is only one microstate for the given parameters. The line $\beta = 0$ converging to $a = 1$ for small $\ephi$ marks the onset of negative temperatures, where the number of available microstates starts to decrease with increasing $a$. The parameters of NFW haloes are all assigned negative temperatures. Interestingly the NFW haloes follow rather closely the line of constant $\beta \approx -0.17$, the dotted line.  
\newline
\newline
As discussed in \ref{sectionmicro}, the only modification to the case of unconstrained $\ephi$ is the restriction of the range of $p(z)$  to $0\le z \le 1/\ephi$, and the corresponding change in its quantile function $Q$. While an exact but lengthy and not very insightful expression for $z(x) = Q(1-x)$ can still be written in terms of the error function for any value of $\lambda$ and $\beta$, we prefer to give a characterization through the differential equations
\beq
z'(x) = \exp\lp\ln Z_1[\lambda,\beta] + \lambda z(x) + \beta z^2(x) \rp = u(x)x,
\enq
and
\beq
z''(x) = \lp z'(x) \rp^2 \lp \lambda + 2\beta z(x) \rp.
\enq
The boundary conditions are
\beq
z(1) = 0, \quad z(0) = \frac 1 \ephi. 
\enq
\begin{figure}
\begin{center}
\includegraphics[width = 0.45\textwidth]{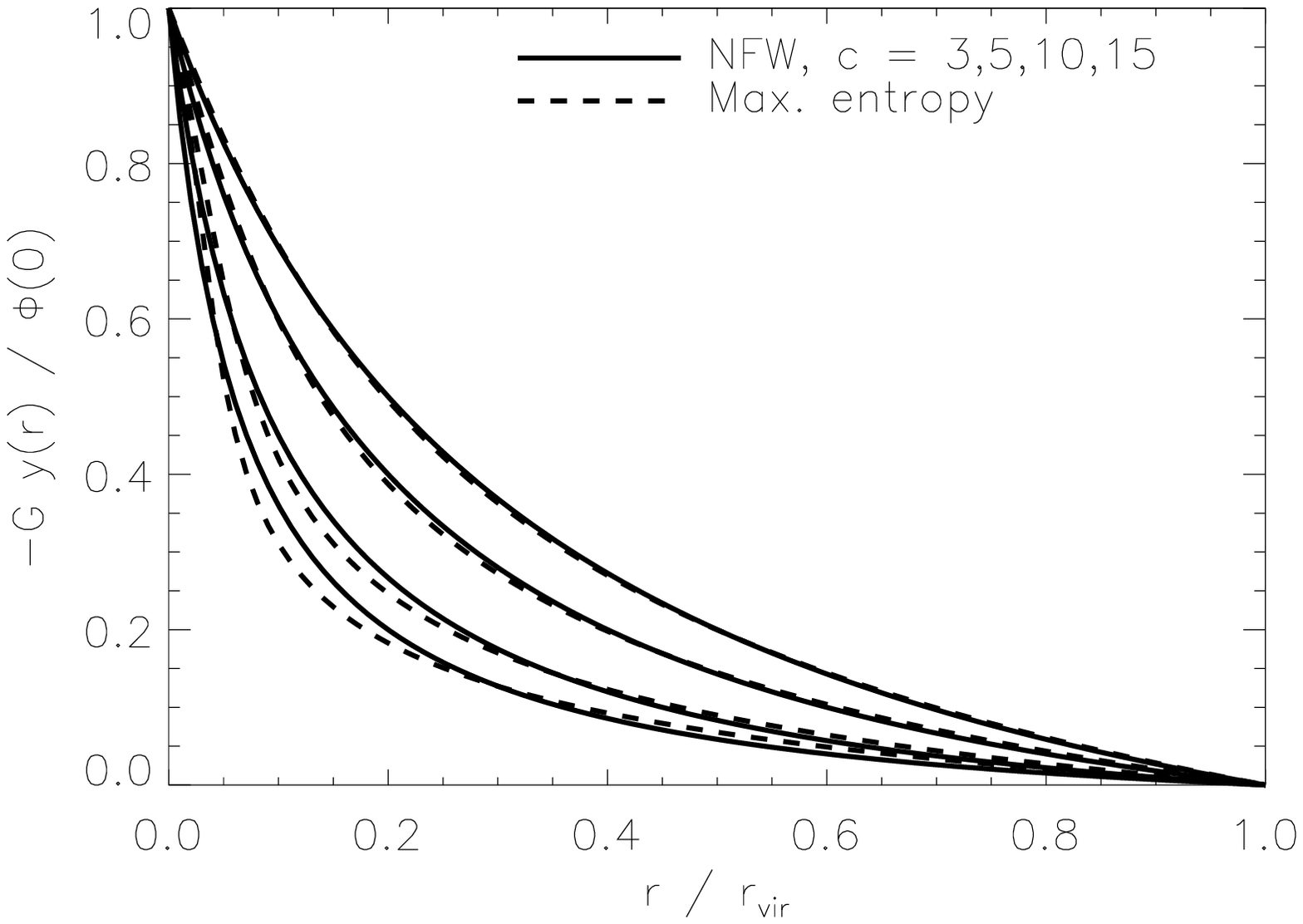}
\includegraphics[width = 0.45\textwidth]{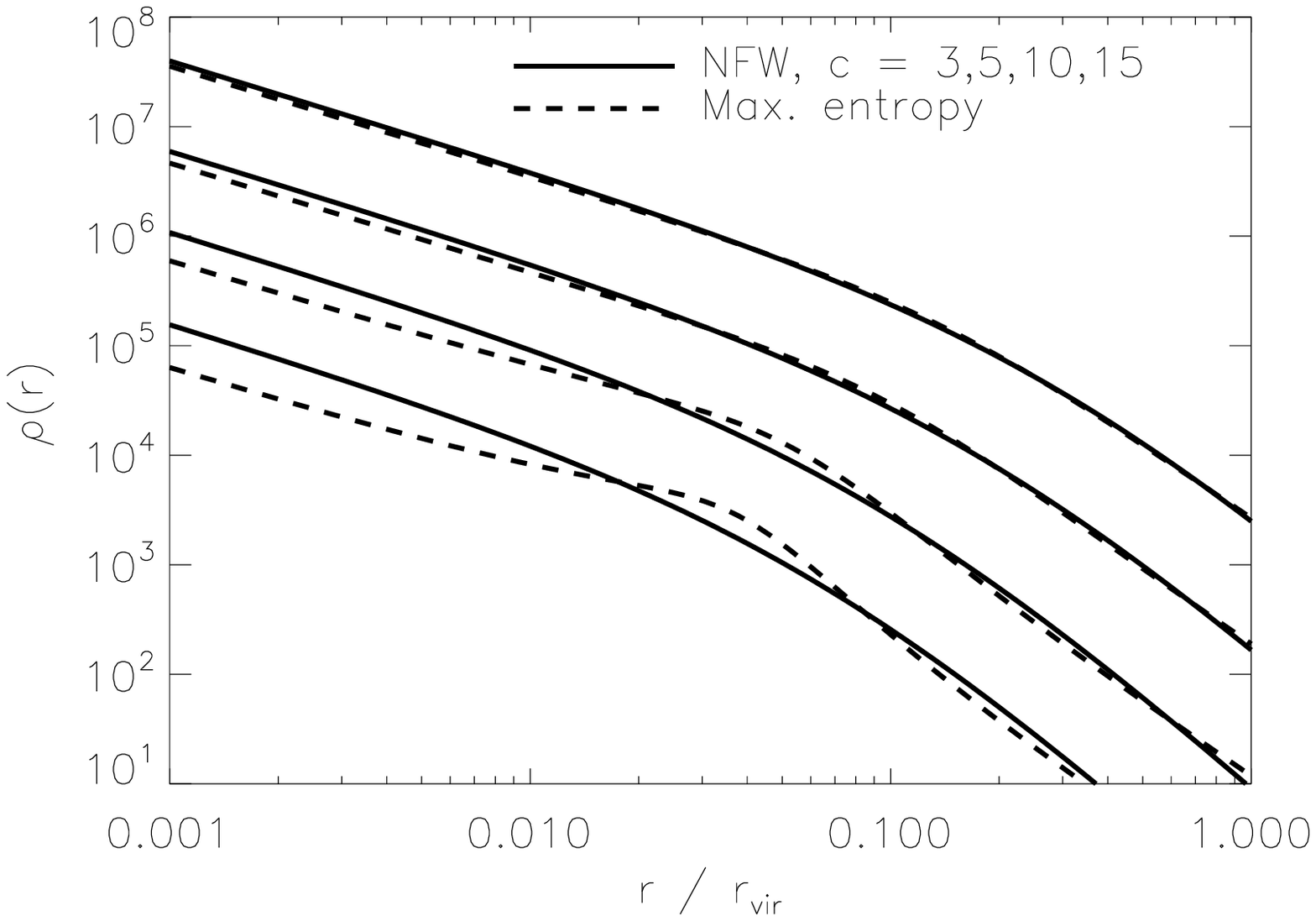}
\caption{Upper panel. The solid lines show the $y$-functions of the NFW profiles, for $c = 3,5,10,15$, from top to bottom. The dashed line the $y$-function singled out by the grand canonical ensemble at the same parameters $a$ and $\ephi$. Lower panel : the same for the density profile, on a logarithmic scale with arbitrary normalization. See text for more details.}
\label{figNFWy}
\end{center}
\end{figure}
Figure \ref{figNFWy} compares the $y$-function (upper panel, normalized to unity at $x = 0$) and the density profiles (lower panel, with consistent but otherwise arbitrary normalization) of the NFW profiles and the one resulting from the statistical ensemble for the same $a$ and $\ephi$. The agreement is very good both qualitatively and quantitatively for massive haloes. At high concentration, too much mass is assigned to the transition region, and the outer slope is not well reproduced anymore. Interestingly, if $\ephi$ is progressively pushed to zero from this point, the transition region is pushed back to $x \approx 0$, the mass assigned to the transition region becomes the additional singularity in the center discussed in \ref{sectionmicro}, and the outer slope becomes uniformly $-2$, the isothermal sphere. The apparition in our ensembles of a characteristic scale radius, roughly $x = 1/c$ for the NFW profiles, is a direct consequence of the finite value forced upon the central potential.
\newline
 
\section{Summary and discussion}\label{sectionconclusion}
In this paper we defined statistical ensembles of virialized spherical halo density profiles at a given mass and gravitational potential energy.  More specifically, we first discretized the relevant radial functions of radius on a regular grid and then obtained their exact thermodynamics in the limit of vanishing cell size. 
 We obtained the mean density profile predicted by theses ensembles, and showed that the ensembles make in fact a very sharp prediction on the $y$-function of the most likely profiles under these constraints. The profiles we found in that way all have a infinitely deep potential in the center, and we recovered the well known singular isothermal sphere in the case where the virial radius is equal to the gravitational radius. Systems with larger ratio show the interesting phenomena of localization of the additional energy in the very center.
\newline
\newline
In essence, our approach is after discretization the statistical mechanics of a gravitationally interacting lattice system, where sites (in fact, radial shells) are fixed and the degrees of freedom are the amount of mass on each shell. This approach differs conceptually both from the statistical mechanics of point particles, where the mass of each particle is fixed but the phase space coordinates of each particle are degrees of freedom, or the approach pioneered by LB discussed in the introduction where the entropy of the one particle phase space distribution only is maximized. A first extension of our approach including velocities would be desirable, though likely difficult, in order to be able to draw additional conclusions and explore deeper connections to these perspectives. 
\newline
\newline
Another difference with respect to traditional statistical mechanics is our restriction to spherically symmetric density profiles. This is justified in our case since we aimed at a description of spherically symmetric haloes only. Nonetheless, it would be interesting  as well to investigate the likely profiles without this constraint. Certainly the preferred profiles would still show this symmetry, but they are likely to differ from those presented in this work. For instance, the profile corresponding to $\beta = 0$ is now a sphere of uniform density, which is never singled out by the ensembles of this work. In the absence of a full solution we can only speculate, but it seems reasonable to expect that for $\beta < 0$ and without further constraints condensation of the additional energy with respect to the case $\beta = 0$ still occurs, but now at an arbitrary point in the volume.
\newline
\newline
Next we connected our results to the NFW density profiles of cosmological dark matter haloes. It is plausible, and has already been suggested \citep{2003MNRAS.339...12Z,2006MNRAS.368.1931L}, that the growth of the central potential is prevented, or is very slow, for cosmological haloes. The central potential emerges in our framework naturally as a fundamental parameter, and implementing this constraint in our ensembles leads to the prediction of a central logarithmic slope of $-1$, as for the NFW profiles seen in cosmological simulations. This holds true at any mass and energy, so that deviations from this relation, as studied in recent simulations \citep{2010MNRAS.402...21N,2009MNRAS.398L..21S} cannot be accommodated in a simple way in our model. At the corresponding values of mass, energy and depth of the potential well, we showed that the NFW profile is very close to the mean value of the statistically preferred profiles, especially the massive haloes. Due to this additional constraint, these systems are assigned negative temperatures. The number of microstates available is actually decreasing if additional energy $|W|$ is given to the system. This suggests that the NFW profiles might not be the end stage of gravitational evolution, rather, a long lived quasi equilibrium state. 
\newline
\newline
It should be stressed that the statistical arguments presented here do not explain from first principles the emergence of these profiles : an explanation for the location of the NFW haloes in the relevant parameter space shown in figure \ref{figNFWc} is left for future research. This location might be related to the type of initial conditions encountered in cosmological simulations, or to any physical process that our approach cannot accommodate.  Nevertheless, the appearance of the inner slope of unity, and that of a characteristic radius is a common feature of all the profiles singled out by the ensembles at any mass and energy, as long as the central potential is forced to take a finite value. These results possibly shed new light on the respective roles of physics and statistics in the emergence and stability of these profiles.

\section*{Acknowledgments}
The authors acknowledge NASA grants NNX12AF83G and NNX10AD53G for support, and thank Mark Neyrinck, Alex Szalay and Albert Stebbins for useful discussions. We are grateful to the reviewer for useful comments and suggestions.
\begin{appendix}
\section{Mass and energy as first and second moments}
This work relies heavily on the fact that the mass and gravitational potential energy are the first and second integral of the $\y$-function, equations \eqref{mass} and \eqref{energy}, with
\beq \label{def2}
y(r) = 4\pi \int_r^{\rvir} ds\:s\:\rho(s).
\enq
We provide here one of the ways to prove this assertion.
\newline
\newline
Consider first equation \eqref{mass}. Plugging in the above definition of $\y(r)$ and reorganizing the integrals we have
\beq
\begin{split}
\int_0^{\rvir} dr\:\y(r) &= 4\pi \int_0^{\rvir} dr\: \int_r^{\rvir} ds\:s \:\rho(s)\\ &= 4\pi \int_0^{\rvir} ds\:s \:\rho(s) \underbrace{\int_0^s dr}_{s}. 
\end{split}
\enq
This last expression is the mass of the halo and \eqref{mass} is proved. To show \eqref{energy}, we proceed similarly, using the definition \eqref{def2} and performing the $r$-integral, which now leads to
\beq
\begin{split}
&-\frac G 2 \int_0^{\rvir} dr\: y^2(r) \\
&= -\frac G 2 \lp 4\pi\rp^2 \int_0^{\rvir}ds_1 \int_0^{\rvir} ds_2\: s_1\rho(s_1)\:s_2\rho(s_2) \min(s_1,s_2).
\end{split}
\enq
To go further, we use the following trick
\beq
\begin{split}
s_1s_2\: \min(s_1,s_2) &= \frac{s_1^2s_2^2 }{\max(s_1,s_2)}\\& = s_1^2s_2^2 \int_0^\infty dr \:\frac 1 {r^2} \theta(r-s_1)\theta(r-s_2),
\end{split}
\enq
where $\theta(x)$ is the unit step function.
The $s_1$ and $s_2$ integrals form now together with the factor of $\lp4\pi\rp^2$ simply $M^2(\le r)$, and we obtain
\beq
-\frac G 2 \int_0^{\rvir} dr\: y^2(r) = -\frac G 2 \int_0^\infty dr \: \lp \frac{M\lp \le r \rp  }{r}\rp^2
\enq
which was to be shown.
\end{appendix}
\bibliographystyle{mn2e}
\bibliography{bib}
\end{document}